\documentstyle{article}
\title{
\begin{flushright}
{\bf\normalsize   COLO-HEP-324\\LPTHE-93-39}\\ \end{flushright}
\bf 2d O(3) + 2d QG}

\author{ {\it C.F. Baillie} \\
	 Computer Science Dept. \\ University of Colorado\\ Boulder, CO 80309,
	 USA\\ \\
	 and
         \\ \\
         {\it D.A. Johnston}\\
         LPTHE\\
	 Universite Paris Sud, Batiment 211\\
         F-91405 Orsay, France$^{1}$\\
	 \\
         Dept. of Mathematics\\
         Heriot-Watt University\\
         Edinburgh, EH14 4AS, Scotland$^{2}$ }

\begin{document} \maketitle
		      {\Large \begin{abstract}
%
It has been suggested that the peak in the specific heat observed
numerically for random surface actions with extrinsic curvature
on dynamical lattices might be the result of a low mass bound state
in an asymptotically free theory, rather than the signal for a
real phase transition. The $O(3)$ model on a fixed lattice displays
just such behaviour, but in general transitions appear to be
weakened when the models concerned are put on dynamical lattices
(ie coupled to 2d quantum gravity).
We have therefore
performed simulations of the $O(3)$ model on dynamical $\phi^3$ graphs
to see if there is still a peak in the specific heat and
compared the results with those on fixed lattices.
\\ \\
Submitted to Phys Rev D.  \\ \\ \\ \\ \\ \\ \\ \\
$1$ {\it Address Sept. 1993 - 1994} \\
$2$ {\it Permanent Address}
%
			\end{abstract} }
%
  \thispagestyle{empty}
%
%
  \newpage
%
		  \pagenumbering{arabic}

There has been a considerable amount of numerical work
devoted to exploring the issue of whether a non-trivial
continuum limit exists for actions of the form
\begin{equation}
S = \sum_{<ij>} ( \vec X_i - \vec X_j )^2 +
\lambda \sum_{\Delta_i, \Delta_j} ( 1 - \vec n_i  \cdot \vec n_j )
\label{e01}
\end{equation}
on dynamically triangulated random surfaces
where the $\vec n_i$ are the unit normals on neighbouring triangles.
The theory apparently has a low $\lambda$ crumpled phase
and a large $\lambda$ smooth phase similar to those displayed
by similar actions on {\it fixed} triangulated surfaces \cite{0}.
The
dynamical lattice simulations
are of interest for constructing well-defined lattice
versions of string theory \cite{1,2,3,4}. The
initial simulations in \cite{1} saw what appeared to be a second order
transition on small lattices, but later work \cite{2,3,4} with larger
lattices and better statistics has shown that the transition, if it
exists, is most probably higher order. Indeed, it was observed in
\cite{3} that the data was not inconsistent with a crossover as the
correlation length reached the size of the surfaces simulated or with
a low mass bound state in an asymptotically free theory. The further
simulations of \cite{4} tend to exclude the first hypothesis, but the
second remains a possibility.

The original one-loop calculation of Polyakov \cite{5} for
the random surface action with extrinsic
curvature suggested that the extrinsic curvature coupling
was asymptotically free and hence there was no phase transition.
The second (extrinsic curvature) part of the action in equ.(\ref{e01})
is, in effect, a sigma model living on the dynamically
triangulated surface subject to the constraint that the $\vec n$'s
be normals to the surface. If one forgets these constraints
for a surface embedded in
three dimensions, which is where most of the
numerical simulations have been done,
it is a $2d$ $O(3)$ model
\begin{equation}
S = - \lambda \sum_{<ij>}  \vec n_i  \cdot \vec n_j
\label{e02}
\end{equation}
with $|\vec n_i |^2 = 1$, which is known to be asymptotically free.
The constraints on the normals in \cite{5} modified the numerical
coefficient in the one-loop beta function
by comparison with the $O(3)$
model but did not change its sign. A more recent calculation by
Polyakov for a surface embedded in four dimensions with an action
incorporating a topological term has in fact shown that the constraints
on the normal vectors are softened by renormalization and may thus, in
this case, be irrelevant anyway \cite{6}.
Intriguingly, a calculation of the renormalization flow of the coupling has
recently
been carried out for an ``standard'' $O(N)$ model coupled to $2d$ gravity
\cite{6a}
where it was found that the result without gravity was renormalized by a factor
$(2 / Q \alpha)$, where $Q = \sqrt{( 26 - N )/ 3} $
and $\alpha = (\sqrt{26 - N} - \sqrt{ 2 - N}) / \sqrt{12}$.
The rescaling becomes complex for the region
$N > 2$ (ie $C>1$) where the model without gravity
is asymptotically free, which is yet another
manifestation of the $C=1$ barrier in continuum $2d$
quantum gravity. It is thus unclear what to expect from
a simulation of the $O(3)$ model on dynamical lattices
from these analytical arguments.

It was suggested in \cite{4} that the
$2d$ $ O(3)$ model might serve as a qualitative model
for the behaviour of discretized random surface actions with extrinsic
curvature
at the crumpling transition. For a fixed lattice $2d$ $O(3)$ model
one observes a
peak in the specific heat that increases with system size for small
lattices before saturating
even though the model is known to be asymptotically free.
The peak is explained by the excitation of
a low mass state, the $\sigma$ particle \cite{7,8}, which
contributes when its mass is comparable
to the inverse correlation length of the lattice $O(3)$ model.
A similar peak is observed in $SU(2)$ lattice gauge theory,
and it appears in both these cases that there are complex zeroes
lurking near the real axis which do not converge to it in the infinite
volume limit \cite{9}.
In general \cite{10} transitions appear to be
weakened when models are transcribed from
fixed to dynamical lattices so it is a priori possible that
the already weak peak in the specific
heat of the $2d$ $O(3)$ model might get washed out completely when the
model is put on a dynamical lattice thus invalidating the qualitative
parallels that have been drawn between the {\it fixed} lattice $2d$
$O(3)$ model and the dynamically triangulated random surface action of
equ.(\ref{e01}).

In this paper we conduct a simulation of the $2d$ $O(3)$ model
on dynamical $\phi^3$ graphs, which is therefore one step closer
to the simulations of the surface action in equ.(\ref{e01}) than the
fixed lattice $O(3)$ model simulations. In our case we are neglecting
``only'' the gaussian term and the constraints on the vectors $\vec
n_i$. The simulations can also be regarded as a test of whether
anything catastrophic happens to the lattice $O(N)$ model for $N>2$,
as the results of \cite{6a} suggest.
We choose to employ a simple Metropolis update for the
spins, which at first sight might appear rather perverse in view
of the excellent cluster update methods that are now available
\cite{11}. There are two reasons for this choice:
Firstly, the simulations of equ.(\ref{e01})
which we are attempting to approximate employed
local algorithms as cluster updates appear to be very difficult,
if not impossible, to implement for actions
containing extrinsic curvature terms especially on dynamically
triangulated surfaces; Secondly, a simulation
of the $O(3)$ model on dynamical $\phi^3$ graphs using the Wolff
cluster algorithm revealed that there are equilibration
problems between the mesh and the spins on dynamical triangulations
that first appear at quite modest $\beta$ values. This is
peculiar to dynamical lattices, as a simulation of the $O(3)$
model on a fixed $\phi^3$ graph gave identical results for
the Metropolis and Wolff algorithms. Rather counterintuitively
the solution to the problem when using the Wolff algorithm appears to be
the use of {\it less} flip moves per Wolff sweep as $\beta$
increases.

For the measurements
reported here we simulate the $O(3)$  model on dynamical $\phi^3$ graphs
of spherical topology with a fixed number $N$ of points and no tadpole
or self-energy insertions which would correspond to degenerate
triangulations on the direct lattice.
We simulate graphs with
$N$ from $100$ to $10000$ points for a range of
$\beta$ values from 0.05 to 10.0. For each data point we carried out
10,000 metropolis equilibration sweeps, followed by 50,000
metropolis measurement sweeps.
We used the usual rule of thumb of carrying out $N$
flip moves per metropolis sweep, which was sufficient to ensure
equilibration between the lattice and spin model.
We measured the usual spin model properties, the energy
\begin{equation}
E = { 1 \over N} < \sum_{<ij>}  \vec n_i  \cdot \vec n_j >,
\end{equation}
specific heat
\begin{equation}
C = { \lambda^2  N} \left( <E^2> - <E>^2 \right)
\end{equation}
and susceptibility. The metropolis acceptance was also measured during the
course
of the simulation and the proposed moves adjusted so that
it remained in the region of fifty percent as $\beta$ was varied.
For the lattice itself
we measure $AL$ and $AF$ which relate to the acceptance of the
flip moves on the graphs. A flip can be forbidden either from
constraints arising from the graph (ie no tadpoles and no self-energy
bubbles) or from the energy change in the spin model induced by the
reconnection of the vertices. $AL$ measures the fraction of randomly
selected links which pass the first test and could be flipped according
to the graph constraints and $AF$ measures the fraction of the links
satisfying the graph constraints that actually are flipped, ie pass the
Metropolis test using the $O(3)$ model energy change. We also measure
the fraction of rings of length 3, $PR3$, which serves as an indicator
of the local curvature distribution in the $\phi^3$ graph.

Remembering that each point on the
lattice is trivalent and taking into account our summation conventions
we would expect the energy to approach $4/3$ for asymptotically large
$\beta$. We find that this is, indeed, the case for the metropolis
algorithm for both fixed and dynamical $\phi^3$ graphs, although
the approach is rather slower on dynamical graphs and does not reach
$E \simeq 1.33$ until $\beta \simeq O(10)$.
The Wolff algorithm with the number of flips per measurement sweep
taken as $N$ fails to reach this asymptotic value completely.
We are of course
principally interested in the behaviour of the specific heat curve
on dynamical lattices and this is shown in Fig.1 for $\beta$ up
to $4.0$. It is clear from the graph that the specific heat looks very
similar to that of the fixed lattice $O(3)$ model.
The peak in the specific heat thus still mimics rather well the behaviour of
the
dynamically triangulated random surface action in equ.(1). The peak grows
for relatively small numbers of points but saturates with larger $N$ - the
points for $N=10000$ in Fig.1 are rather poorly equilibrated and are probably
best discarded in making comparisons. In addition to showing that
the $O(3)$ model on dynamical lattices still produces
similar qualitative behaviour
to the random surface action the simulation provides another example
of numerical work venturing into regions where continuum analytical
calculations
suggest trouble (in this case the complex scaling found in \cite{6a} for $N>2$)
and finding nothing untoward. Simulations and series extrapolation
of multiple Ising models on dynamical
lattices which also have $C>1$, at least naively, also fail to show any
pathologies \cite{12}, whereas continuum calculations produce
complex critical exponents \cite{13}.

The behaviour of the dynamical $\phi^3$ graphs is shown in Fig.2.
where $AF$, $AL$ and $PR3$ are plotted for $N=5000$. There is a drop in the
flip acceptance $AL$ that is associated with the graph constraints at
the same position as a peak in the probability of rings of length three, $PR3$,
at a $\beta$ value slightly below the peak in the specific heat.
There is also a dip in the actual flip acceptance $AF$ that is slightly shifted
from this. These characteristics are again surprisingly similar to those
observed in earlier random surface simulations, where similar peaks and dips
are also manifest. This behaviour is not totally universal for all models
on dynamical lattices, as $AF$ and $AL$ look rather different for the $XY$ and
$SOS$ models \cite{14}, though Potts and multiple Potts models on dynamical
lattices
which have genuine phase transitions
give results similar to those seen here for a model which presumably does not.

In summary, we have added to the repertoire of models that have been simulated
on
dynamically triangulated lattices or their duals by simulating a model
that is asymptotically free on a fixed lattice. We found that the $O(3)$ model
still possessed a peak in its specific heat that eventually stopped growing
for large $N$ and displayed very similar lattice characteristics
to random surface actions too. These results, like the fixed lattice $O(3)$
results in \cite{4}, show that it is not impossible that the observed
crumpling ``transition'' for fluid random surfaces is a lattice artifact in
an asymptotically free theory.
In addition the results here are another example of a  simulation in the region
beyond the
$C=1$ barrier of the continuum formulation which have found no obvious
pathologies in
either the graphs or the spin model. From the algorithmic point of view
a comparison of the metropolis algorithm used here and a Wolff algorithm
simulation
revealed that there were unexpected equilibration problems
on dynamical lattices for the Wolff algorithm as $\beta$ was increased.
In view of the results of \cite{6a} it would be most interesting to carry
out a careful measurement of scaling on a dynamical lattice close to $\beta=0$
(where the Wolff algorithm is still well behaved) in order to see how the
renormalization group flow compared with the fixed lattice model, and we are
currently
doing this.

This work was supported in part by NATO
collaborative research grant CRG910091.
CFB is supported by DOE under
contract DE-FG02-91ER40672 and by NSF Grand Challenge Applications
Group Grant ASC-9217394
DAJ is supported at LPTHE
by an EEC HCM fellowship, EEC HCM network grant and an Alliance grant.
The simulations were carried out on
workstations at Heriot-Watt University and the
Cray YMP-EL at LPTHE, Orsay.

\vfill \eject  \vfill \eject
\centerline{\bf Figure Captions}
\begin{description}
\item[Fig. 1.] The specific heat
for the various graph sizes simulated using a metropolis
algorithm for the spins.
\item[Fig. 2.] The flip acceptances, $AL$ $AF$, and the probabilities of rings
of length three, $PR3$, for a graph with $N=5000$.
\end{description}

\begin{thebibliography}{99}
\bibitem{0} Y. Kantor and D. Nelson, Phys. Rev. Lett. {\bf 58} (1987)
2774;\\
            J. Ambjorn, B. Durhuus and T. Jonsson, Nucl. Phys. {\bf
B316} (1989) 526;\\
            R. Harnish and J. Wheater, Nucl. Phys. {\bf B350} (1993)
447;\\
            J. Wheater and P. Stephenson, Phys. Lett. {\bf B302} (1993)
447.
\bibitem{1} S. Catterall, Phys. Lett. {\bf 220B} (1989) 207;\\
            C. Baillie, D. Johnston and R. Williams, Nucl. Phys. {\bf
B335} (1990) 469;\\
            C. Baillie, S. Catterall, D. Johnston and R. Williams, Nucl.
Phys. {\bf B348} (1991) 543;\\
            S. Catterall, D. Eisenstein, J. Kogut and R. Renken, Nucl. Phys.
{\bf B366} (1991) 647.
\bibitem{2} J. Ambjorn, A. Irback, J. Jurkiewicz and B. Petersson, Nucl.
Phys. {\bf B393} (1992) 571;\\
            J. Ambjorn, A. Irback, S. Varsted, J. Jurkiewicz and B. Petersson,
Phys. Lett. {\bf 275B} (1992) 295.
\bibitem{3} M. Bowick, P. Coddington, L. Han, G. Harris and E. Marinari,
Nucl. Phys. {\bf B394} (1993) 791.
\bibitem{4} K. Anagnostopoulos, M. Bowick, P. Coddington, L. Han,
            G. Harris and E. Marinari, ``Fluid Random Surfaces with Extrinsic
Curvature: II'',
SU-HEP-93-4241-540, hep-lat/9308091.
\bibitem{5} A. Polyakov, Nucl. Phys. {\bf B268} (1986) 406.
\bibitem{6} A. Polyakov, Princeton University Preprint, PUPT-1394.
\bibitem{6a} I. Klebanov, I. Kogan and A. Polyakov, ``Gravitational Dressing of
            the Renormalization Group'', Princeton University Preprint,
PUPT-1421, hep-ph 9309106.
\bibitem{7} G. Martinelli, G. Parisi and R. Petronzio, Phys. Lett.
            {\bf 100B} (1981) 485;\\
            J. L. Colot, J. Phys. {\bf A16} (1983) 4423.
\bibitem{8} R. Brout and W. Deans, Nucl. Phys. {\bf B215} (1983) 407;\\
            R. Brout, W. Deans and A. Silovy, Phys. Rev. {\bf B27}
(1983) 5813;\\
            R. Brout and J. Orloff, Nucl. Phys. {\bf B270} (1986) 273.
\bibitem{9} P. Butera, M. Comi and G. Marchesinin, Nucl. Phys. {\bf
B300} (1989) 1;\\
            M. Falcioini, E. Marinari, M. Paciello, G. Parisi and B.
Taglienti, Phys. Lett. {\bf 102B} (1981) 270; Nucl. Phys. {\bf B190}
(1981) 782; Phys. Lett. {\bf 10b} (1982) 331;\\
           E. Marinari, Nucl. Phys. {\bf B235} (1984) 123.
\bibitem{10} For a review see ``Matrix Models of 2d Gravity'' by P.
Ginsparg, Trieste Lectures 1991,
	    LA-UR-91-9999, hep-th 9112013.
\bibitem{11} U. Wolff, Nucl. Phys. {\bf B334} (1990) 581.
\bibitem{12} C. F. Baillie and D. A. Johnston, Phys. Lett. {\bf B286}
            (1992) 44;\\ J. Ambj\o rn, B. Durhuus and T. Jonsson,
            Nucl. Phys. {\bf B398} (1993) 568;\\
             S.M. Catterall, J.B. Kogut and R.L. Renken, Phys. Lett.
             {\bf B292} (1992) 277;\\ S. Hikami, Phys. Lett. {\bf B305}
            (1993) 327;\\ E. Brezin and S. Hikami, Phys. Lett. {\bf
            B295} (1992) 209; Phys. Lett. {\bf B283} (1992) 203
\bibitem{13} V.G. Knizhnik, A.M. Polyakov and A.B. Zamolodchikov, Mod.
Phys. Lett. {\bf A3} (1988) 819;\\F. David, Mod. Phys.
Lett. {\bf A3} (1988) 1651;\\
            J. Distler and H. Kawai, Nucl. Phys. {\bf B321} (1989)
            509.
\bibitem{14} C. Baillie and D. Johnston, Phys. Lett. {\bf B291} (1992) 233;\\
C. Baillie, W. Janke and
D. Johnston, ``DGSOS on DTRS'', COLO-HEP-320, LPTHE 93/34.
\end{thebibliography}
\end{document}